\documentclass[journal,12pt,onecolumn,draftclsnofoot,]{IEEEtran}
\ifCLASSOPTIONcompsoc
\usepackage{placeins}
\usepackage{lipsum}

\usepackage{floatrow}
\usepackage{float}
\restylefloat{table}
  \usepackage[nocompress]{cite}
\else
  \usepackage{cite}
\fi
\usepackage{booktabs}
 \usepackage{graphicx}
 \usepackage{subfigure}
 \usepackage{xspace}
\usepackage{enumitem}

 \usepackage{amsmath,amssymb,amsthm,amscd,amstext}
 \usepackage{color}
\AtBeginDocument{%
  \addtolength\abovedisplayskip{-0.3\baselineskip}%
  \addtolength\belowdisplayskip{-0.3\baselineskip}
}
\makeatletter
\newcommand\xleftrightarrow[2][]{%
\ext@arrow 9999{\longleftrightarrowfill@}{#1}{#2}}
\newcommand\longleftrightarrowfill@{%
\arrowfill@\leftarrow\relbar\rightarrow}
\makeatother

\usepackage{soul}

\newcommand{\sol}{{LightIoT}\xspace }

\makeatletter
\def\ps@IEEEtitlepagestyle{%
  \def\@oddfoot{\mycopyrightnotice}%
  \def\@evenfoot{}%
}
\def\mycopyrightnotice{%
  {\footnotesize  \textcolor{red} {This paper has been accepted for publication by the IEEE Transactions on Green Communications and Networking. The copyright is with the IEEE.} \hfill}
  \gdef\mycopyrightnotice{}
}

\begin{document}

\title{\sol: Lightweight and secure communication for energy-efficient IoT in health informatics }


\author{Mian~Ahmad~Jan,~\IEEEmembership{Senior Member~IEEE,}
        	Fazlullah~Khan$^*$,~\IEEEmembership{Senior Member~IEEE,}
        	Spyridon Mastorakis$^*$,~\IEEEmembership{Member~IEEE,}
        	Muhammad Adil,~\IEEEmembership{Student Member~IEEE,}
        	Aamir Akbar,~\IEEEmembership{Member~IEEE,}
        	and Nicholas Stergiou.
     
\thanks{A * indicates the corresponding authors.}
\thanks{Mian Ahmad Jan, Fazlullah Khan, and Aamir Akbar are with the Department of Computer Science, Abdul Wali Khan University Mardan, 23200, KPK, Pakistan. (E-mail: mianjan@awkum.edu.pk, fazlullah@awkum.edu.pk, amirakbar@awkum.edu.pk)
}
\thanks{Spyridon Mastorakis is with the Department of Computer Science, University of Nebraska at Omaha, NE, USA 68182-0002 (E-mail: smastorakis@unomaha.edu)
}
\thanks{Muhammad Adil is with the Department of Computer Science, Virtual University of Pakistan, 54–Lawrence Road, Lahore 54000. E-mail:muhammad.adil@ieee.org}
\thanks{Nicholas Stergiou is with the Department of Biomechanics, University of Nebraska at Omaha, NE, USA 68182-0002 (E-mail: nstergiou@unomaha.edu)
}
}

\maketitle

\begin{abstract}
Internet of Things (IoT) is considered as a key enabler of health informatics. IoT-enabled devices are used for in-hospital and in-home patient monitoring to collect and transfer biomedical data pertaining to blood pressure, electrocardiography (ECG), blood sugar levels, body temperature, etc. Among these devices, wearables have found their presence in a wide range of healthcare applications. 
These devices generate data in real-time and transmit them to nearby gateways and remote servers for processing and visualization. The data transmitted by these devices are vulnerable to a range of adversarial threats, and as such, privacy and integrity need to be preserved. In this paper, we present \sol, a lightweight and secure communication approach for data exchanged among the devices of a healthcare infrastructure. \sol operates in three phases: initialization, pairing, and authentication. These phases ensure the reliable transmission of data by establishing secure sessions among the communicating entities (wearables, gateways and a remote server). \textcolor{black}{ Statistical results exhibit that our scheme is lightweight, robust, and resilient against a wide range of adversarial attacks and incurs much lower computational and communication overhead for the transmitted data in the presence of existing approaches. }
\end{abstract}



\section{Introduction}

\IEEEPARstart{T}{he} 21$^{st}$ century has witnessed significant advancement in the development of smart devices and wireless communication technologies.  
These devices and technologies have found their presence in numerous applications such as smart healthcare, smart industrial automation, and smart surveillance~\cite{jan2020lightweight, rehman2020ccic, jan2020security}. 
In these applications, the Internet of Things (IoT) interconnect various sensors, actuators and smart devices  with the edge servers and cloud data centres by regulating the exchange of data among them~\cite{mastorakis2020icedge}. In the context of healthcare, the IoT is assumed to connect medical devices with the communication technologies to enable new applications by supporting intelligent decision-making for healthcare data~\cite{al2018context}. Like IoT, smart healthcare technologies have improved at a rapid pace due to a massive increase in the volume of biomedical data.  
Therefore, IoT can play a pivotal role in the development of cost-effective and smarter healthcare applications that can monitor the patients in real-time to save their lives in an event of emergency, e.g., heart failure, sudden and acute pain, asthma attack, etc. The proliferation in mobile communication bridges the gap among the smart devices and the practitioners by providing seamless and reliable delivery of gathered data~\cite{al2018context, jan2021ai}. This proliferation has led to a patient-centric approach that enables the remote monitoring of patients with shorter hospital stays and, in most cases, avoiding the hospital altogether.  

Healthcare devices such as smart watches, fitness trackers, etc. have enabled improvements in quality of living in recent years~\cite{fitbit, bader2019internet}. 
These devices sense human activities and generate real-time data about step count, sleep cycle, heart rate, and pulse count, breathing rate, and others.
These devices are typically low-powered, resource-constrained, and transmit the gathered data to a nearby mobile device using wireless communication technologies~\cite{banerjee2018wearable}. For these devices, green communication is highly desirable to conserve their resources. The biomedical data generated by these devices are always sensitive, confidential, and need to be securely transmitted with their privacy preserved. In wireless networking, the communication channels are lossy and prone to various malicious attacks, for example, Denial of Service (DoS), Sybil, impersonation, and eavesdropping, are a few to mention. To this end, smart healthcare devices need to be secure, tamper-proof, and accessed by authorized and authentic users only.  In health informatics, unauthorized access to the data by adversaries can wreak havoc in the healthcare sector~\cite{mastorakis2021dlwiot}. 

In traditional communication networks, data security techniques are strong enough to defend against various adversarial attacks. 
These techniques are based on cryptography and provide data security and privacy at the expense of network resources. However, IoT-enabled smart healthcare systems have different requirements in terms of data security and system architecture. In this context, the existing cryptography-based solutions cannot be migrated directly~\cite{zhang2017security}. 
In a smart healthcare system, the devices are connected with the Internet via a gateway that expose them to various malevolent entities. 
If these devices are compromised, it will be difficult to predict the nature of attacks posed by them. 
As a result, smart healthcare applications face botnets along with thingbots at the same time \cite{jan2019payload}. 
To secure the system, data integrity, data confidentiality, data availability, authenticity, and non-repudiation need to be considered~\cite{he,khan2019secured, khan2020secured}. 
Therefore, it is essential to address these challenges while keeping the resource-starving nature of healthcare devices in mind.

In this paper, we propose \sol, a lightweight approach for the secure transmission of biomedical data among the communicating entities. \textcolor{black}{\sol provides secured exchange of data and robust registration for devices interested in communication. Our approach is equally applicable in any application of IoT that has security requirements, e.g., industrial automation, smart homes, smart cities,  etc. } In \sol, the resource-constrained wearables collect patients data and transmit them to a remote server via gateways (mobile terminals)~\cite{mastorakis2020dapes}. \sol operates in three phases and makes the following contributions:

\begin{enumerate} [wide, labelwidth=!, labelindent=0pt]
\item \textcolor{black}{We propose a registration phase for the resource-constrained wearable devices. To avoid excessive delay and computational power, these devices are registered directly with a remote server in an offline phase. They are no longer required to register with the remote server via the intermediate entities, i.e., gateways. The direct registration enables these wearables to immediately deliver time-critical and delay-sensitive biomedical data for decision-making. Besides, this phase ensures green communication by conserving the resources of these wearable devices.}

\item Unlike the existing approaches, we propose a significantly lightweight authentication phase that requires fewer hash functions and Exclusive OR (XOR) operations. Our proposed authentication is lightweight yet highly robust to ensure that secure sessions are established among the communicating devices, i.e., wearables, gateways and a remote server. The presence of non-reproducible pseudo random numbers ensures the privacy preservation of transmitted biomedical data. 
 
 \item Our proposed approach conserves the energy of wearables by prolonging the lifetime of the network. Besides, the lightweight security primitives reduce the end-to-end delay for exchanged messages among the communicating entities. 
\end{enumerate}

The rest of this paper is organized as follow. In Section \ref{related}, the related work pertaining to \sol is presented. \textcolor{black}{In Section \ref{proposed}, the network model of LightIoT is discussed along with its design.} In Section \ref{section5}, we present a detailed security analysis of various malicious threats and the efficiency of \sol in combating them.
In Section \ref{simulation}, we present and validate our experimental results. Finally, the paper is concluded and future research directions are presented in Section \ref{conclusion}.

\section{Related Work} \label{related}
One of the first lightweight user authentication protocols for resource-constrained devices was proposed in 2006 \cite{wong2006dynamic}. This protocol is based on simple operations, such as an one-way hash function and XOR operations. However, this protocol is prone to replay, forgery, and stolen-verifier attacks. In \cite{liu}, a secured healthcare system was proposed using a Wireless Body Area Network (WBAN). 
The use of cryptograhic primitives enable the proposed system to achieve efficiency and robustness and, at the same time, provides transmission confidentiality and authentication among the wearables and a backend server. However, the use of an asymmetric key algorithm, i.e., Elliptic-curve cryptography (ECC), incurs additional overhead for these intelligent wearables. 
In \cite{das2018design}, the authors presented an authentication approach for wearables of a healthcare system. 
The proposed approach allows a user to authenticate his/her wearable device(s) and a mobile terminal, before establishing a session key between them. The use of bitwise XOR operations and hash functions make the proposed approach significantly lightweight for the resource-constrained wearables.
A robust authentication protocol was proposed for intelligent wearables in \cite{diez2015toward}.
This protocol ensures mutual authentication between a wearable and a remote server via the exchange of a session key. This exchange establishes a secure communication channel via the Internet for seamless transmission of biomedical data. 
However, the proposed protocol incurs computation burden on wearables due to the execution of resource-intensive cryptographic primitives.

In \cite{wu2017lightweight}, the authors proposed a lightweight authentication approach with privacy preservation. In this approach, wearables and smartphones mutually authenticate each other in a three-step process by maintaining the anonymity of wearables. The proposed approach used XOR, concatenation, and hash functions for authentication, however, it lacks a clear explanation for achieving anonymity. Besides, it is vulnerable to Sybil, DoS and replay attacks.
In \cite{chang2015provably}, the authors proposed a lightweight and smart card-based authenticated key exchange scheme for resource-constrained devices. 
The proposed scheme uses two authentication protocols to preserves data privacy between resource-constrained devices and a gateway. 
The first protocol uses XOR operations and hash functions, while the second protocol uses elliptic curve cryptography along with XOR and hash functions for authentication. 
\textcolor{black}{ Chang et al.  \cite{chang2015provably} scheme was    investigated in \cite{das2016provably} to verify the effectiveness and vulnerabilities factors. During statistical analysis it has been observed that the proposed model is susceptible to  spoofing and user anonymity attacks.\\
In \cite{he2018security}, the authors investigated the limitations and vulnerabilities of \cite{chang2015provably,das2016provably} by demonstrating an adversary attack on these schemes.  
In  \cite{aghili2019seclap}, the author analyzed the lightweight RFID mutual authentication protocols to resolve the authentication problem in healthcare IoT networks. 
Turkanovic et al. \cite{turkanovic2014novel}, proposed a hash function-based lightweight authentication protocol for wearables healthcare IoT devices to resolve the validation problem in these networks. This protocol was efficient against a wide range of malicious attacks and is capable of authenticating a new device upon joining the network. 
However, this protocol was analyzed by \cite{amin2016secure} and proved that it is vulnerable to impersonation, offline dictionary, and  password-guessing attacks. 
In \cite{amin2016robust}, the authors suggested a lightweight authentication protocol for wearables healthcare IoT devices, which is capable of achieving user anonymity and untraceability by using a dynamic update mechanism. However, this protocol was investigated by  \cite{jiang2017efficient} and proved that it is prone to de-synchronization and cloning attacks. 
This scheme  \cite{amin2016robust}  is not capable to counter forward secrecy and DoS attacks due to  message replacement.}
Recently some lightweight authentication schemes have been proposed for resource-constrained networks \cite{ali2018secure},\cite{li2018robust}, \cite{wu2018lightweight}. Contrary to their claims of being lightweight, these schemes involve too many security primitives during hashing that ultimately make them resource-intensive.   

\section{Network Model and Design of \sol} 
\label{proposed}

\sol consists of three phases: initialization, pairing, and authentication. During the first phase, a remote server generates the system parameters and stores important information about the clients and gateways. The second phase is responsible for registering each client with a trusted server.
Finally, during the third phase,  the clients and gateways mutually authenticate each other via the server, and a session key is generated for the current session to securely exchange the data~\cite{mastorakis2019peer}. In this section, first we discuss the network model of \sol followed by its design. The notations used in each phase of the design are illustrated in Table \ref{tab1}.
\begin{table}[ht]
\begin{center}
\caption{Notations of the \sol Design} \label{tab1}
\begin{small}
\resizebox{0.52\columnwidth}{!}{
\textcolor{black}{
\begin{tabular}{l l}
\textbf{Notations}&\textbf{Descriptions}  \\
\hline \hline  \textit{C} & Client  \\
 GW & Gateway  \\
 S & Server \\
 ID$_C$ & Identity of C \\
 $P_{ID_{c}}$ & Pseudo-identity of C \\
 $\lambda_C$ & Secret key of C  \\
 ID$_{GW}$ & Identity of GW   \\
 $P_{ID_{GW}}$ & Pseudo-identity of GW \\
 $\lambda_{GW}$ &Secret key of GW \\
 $t_{c_1}$, $t_{c_2}$ & Time stamps of C for pairing   \\
 t$_s$ & Time stamp of S for pairing  \\
 r$_c$ & Random Number generated by C for pairing  \\
 $T_{c_1}$, $T_{c_2}$ & Time stamps of C for authentication  \\
 T$_{gw_1}$, T$_{gw_2}$ & Time stamps of GW for authentication \\
 T$_s$ & Time stamp of S for authentication \\
 $\delta T$ & Legal delay time interval \\
 R$_{c}$ & Random Number generated by C for authentication \\
 R$_{gw}$ & Random Number generated by GW for authentication \\
 M$_{1}, M_2, ..., M_6$ & Messages \\
\hline \hline
\end{tabular}
}
}
\end{small}
\end{center}
\end{table}

\subsection{\textcolor{black}{Network Model}} \label{model}

\begin{figure*}[ht]
\centering
\includegraphics[width=5.7in, height=2.3in]{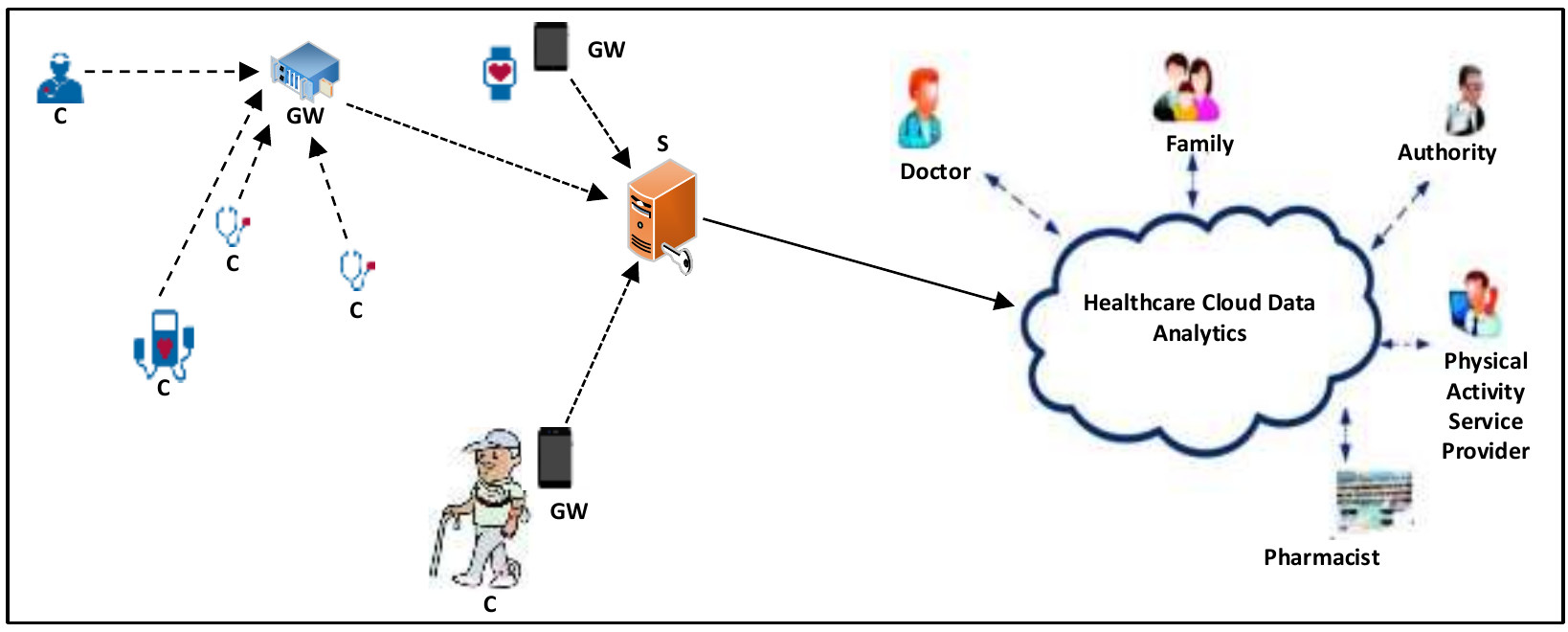}
\caption{Network Model of \sol}
\label{fig:1network}
\end{figure*}

In this section, we discuss our proposed network model that is equally applicable for in-home and in-hospital scenarios. In Fig. \ref{fig:1network}, the sensor-embedded wearables, i.e., clients, are connected to a remote server via the network gateways. For some clients, such as a smartwatch, a smartphone in the patient's pocket acts as a gateway~\cite{yang2020centralized}. These gateways, act as intermediate entities to the remote server that is connected to healthcare cloud data analytics for feature extraction, visualization, and decision-making~\cite{9166765}.

The network model of Fig. 1 is susceptible to various adversarial attacks that can ultimately lead to loss of the associated invaluable medical data~\cite{zhang2020safecity}. An adversary may establish secured connections to the server or gateways if its authentication requests are accepted. An adversary may infiltrate the network by seizing the identities of clients and gateways to pose various threats. Moreover, it may clone itself for a large-scale adversarial effect on the overall system. To prevent such threats, we propose a lightweight yet secure and robust privacy-preserved approach for biomedical data. 
\sol is resilient against the following threats:

\begin{enumerate} [wide, labelwidth=!, labelindent=0pt]
\item Replay: An adversary may replay a stream of previously transmitted messages to the clients or servers.
\item Forgery: An adversary may launch a forgery attack on one or more of the network entities. It may seize and manipulate the exchanged messages and impersonate itself to these legitimate entities. 

\item Anonymity and Untraceability: An adversary may launch this attack by extracting the pseudo-random numbers, and the identities of clients, gateways, and servers from the exchanged messages~\cite{li2019distributed}. In doing so, it may interlink various sessions to maliciously affect these network entities.
\item De-Synchronization: An adversary may launch this attack by blocking the exchanged messages among the communicating entities to alter their sequence/pattern.

\item Key Compromise: An adversary may launch this attack by forging or compromising the exchanged session key.


\end{enumerate}

\subsection{Design of \sol}
The design of \sol consists of three phases: initialization, pairing and authentication. In this section, we discuss these three phases.

\subsubsection{Initialization}

A remote server (\textit{S}) serves the purpose of a trusted third party and generates the systparameters. \textit{S} stores information about each WBAN client (\textit{C}) and each  gateway (\textit{GW}). During initialization, \textit{S} performs the following operations:

\begin{itemize} [wide, labelwidth=!, labelindent=0pt]
\item It provides and stores information, i.e.,  $\lambda_C$ and $P_{ID_{c}}$ for a given \textit{C} in its database as ($\lambda_C$, $P_{ID_c}$). Each \textit{C} has a tuple (\textit{ID$_C$}, $\lambda_C$, $P_{ID_c}$).

\item It provides and  stores information, i.e., $\lambda_{GW}$ and $P_{ID_{GW}}$ for a given \textit{GW} in its database as ($\lambda_{GW}$, $P_{ID_{GW}}$). Each \textit{GW} has a tuple (\textit{ID$_{GW}$}, $\lambda_{GW}$, $P_{ID_{GW}}$) and assumed to be registered and authorized by \textit{S}.

\end{itemize}
\subsubsection{Pairing}

\textcolor{black}{Each \textit{C} needs to register itself with \textit{S} to initiate communication request in the network. During this phase, the following steps are performed.}
\begin{itemize}[wide, labelwidth=!, labelindent=0pt]
\item \textcolor{black}{\textit{C} generates a random number \textit{r$_c$}, picks its current time stamp  \textit{$t_{c_1}$} and calculates the hash \textit{$D_1$} by concatenating $P_{ID_{c}}$, \textit{r$_c$}, and $\lambda_C$, as shown in Eq. \ref{eq1}.  At this point, a message \textit{$M_1$}= \textit{$\{(\textit{ID$_C$}, r_c,  t_{c_1}, D_1 $}\}) is created, XOR with  $P_{ID_{c}}$, i.e., \textit{M$_1$}$ \oplus$ $P_{ID_{c}}$, and the encrypted message is send to \textit{S}}.


\begin{equation}
	D_1 \leftarrow h(P_{ID_c} ||r_c||\lambda_C).
	\label{eq1}
\end{equation}
\item Upon receiving  \textit{$M_1$}, \textit{S} picks up its current timestamp \textit{$t_s$} and checks if $|t_s - t_{c_1}| < \delta T$. If the time interval is not within the specified allowable $\delta T$, then $M_1$ is discarded and pairing request fails, otherwise further processing is performed~\cite{9285216}.
Once \textit{$M_1$} is validated for time, then \textit{S} uses  $P_{ID_{c}}$ to decrypt it for the retrieval of $ID_{C}$. Upon retrieval, \textit{S} checks $ID_{C}$ in its database. If it is found, then it means that \textit{C} is a registered client. Next \textit{S} checks a tuple (\textit{$P_{ID_{c}}$}, $\lambda_C$) in its database, calculates a hash function \textit{$D_1^{'}$} and checks if it  matches the \textit{$D_1$} received from \textit{C}.
If there is a match, it means that the registration/pairing request was received from a legitimate client.
 At this point, \textit{S} calculates a new pseudo-identity $P_{ID_{c}}^{new}$ for \textit{C} by generating the hash of \textit{ID$_C$}, $r_c$, $t_{c_1}$,  and $t_s$, as shown in Eq. \ref{eq2a}.  Here, $P_{ID_{c}}^{new}$ serves as a new pseudo-identity for \textit{C}. Next, \textit{S} generates a hash $D_2$ by encrypting $P_{ID_{c}}^{new}$ with \textit{ID$_{GW}$}, as shown in Eq. \ref{eq2b}. 
Finally, \textit{S} creates a message $M_2$=\textit{$\{D_2, t_s \}$} and broadcasts to \textit{C}.

\begin{subequations}
\begin{equation}
  P_{ID_{c}}^{new} \leftarrow h(ID_C ||r_c||t_{c_1}||t_s), 
\label{eq2a} 
\end{equation}    
\begin{equation}
 D_2 \leftarrow h(P_{ID_{c}}^{new} \oplus ID_{GW}).
 \label{eq2b}
\end{equation}
\end{subequations}

\item Upon receiving $M_2$, \textit{C} picks up its current timestamp $t_{c_2}$ and checks if $| t_{c_2}- t_s| < \delta T$. If the time interval is not within the specified allowable $\delta T$, then $M_2$ is discarded, otherwise, further processing is performed. 
After the validation of $M_2$, \textit{C} calculates its new pseudo-identity $P_{ID_{c}}^{new}$  by generating the hash of $ID_C$,  $r_c$,  $t_{c_1}$  and $t_s$, as shown in Eq. \ref{eq3a}. In this case, $P_{ID_{c}}^{new}$  serves as the new pseudo-identity for \textit{C}. It is worth mentioning that the same pseudo-identity for \textit{C} was earlier generated by \textit{S}. \textit{C} then obtains \textit{ID$_{GW}$} from $D_2$ using Eq. \ref{eq3b}, calculates \textit{$D_2^{'}$}, and checks whether \textit{$D_2^{'}$} matches $D_2$. If a correct hash function is calculated, then the session request for pairing is validated, and \textit{C} updates its pseudo-identity, i.e., $P_{ID_{c}}^{new}$ becomes the new $P_{ID_{c}}$. The complete procedure of our pairing phase is shown in Fig. \ref{fig1}.

\begin{figure*}
\centering
  \includegraphics[width=15cm, height=6cm]{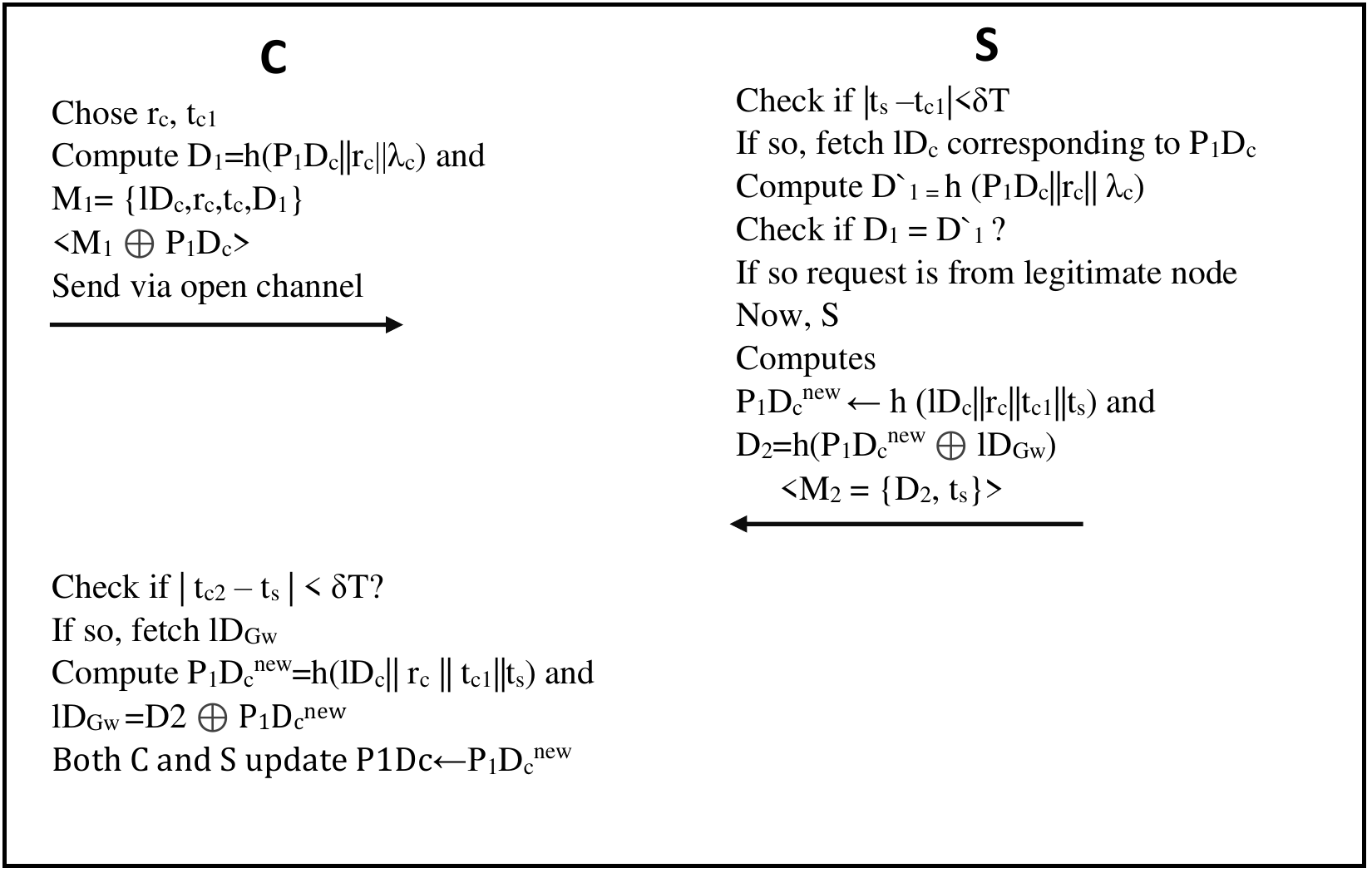}
  \caption{Pairing Phase}
  \label{fig1}
\end{figure*}

%

\begin{subequations}
\begin{equation}
P_{ID_{c}} \leftarrow P_{ID_{c}}^{new} \leftarrow h(ID_C ||r_c||t_{c_1}||t_s),
\label{eq3a}
\end{equation}    
\begin{equation}
ID_{GW} \leftarrow D_2 \oplus h(P_{ID_{c}}^{new}||t_s).
 \label{eq3b}
\end{equation}
\end{subequations}

\end{itemize}

\subsubsection{Authentication}

\textcolor{black}{For authentication, all the three entities (\textit{C}, \textit{GW}, and \textit{S}) participate. During this process, \textit{C} and \textit{GW} mutually authenticate each other and a session key is generated for further communication between them. The following steps are involved during the authentication phase.}

\begin{enumerate} [wide, labelwidth=!, labelindent=0pt]

\item \textit{C} generates a random number \textit{R$_c$}, picks up its current timestamp \textit{$T_{c_1}$}, and calculates a hash \textit{C$_1$} by concatenating \textit{ID$_C$}, $\lambda_C$, and \textit{R$_c$}, as shown in Eq. \ref{eq6}. At this point, a message $M_3$=\textit{$\{C_1, R_c, T_{c_1}, P_{ID_c} \}$} is created, XOR with $ID_{GW}$, i.e., \textit{M$_3$}$ \oplus$ $ID_{GW}$, and the encrypted message is sent to \textit{GW}, as an authentication request. 

\begin{equation}
	C_1 \leftarrow h(ID_C ||\lambda_C||R_c).
	\label{eq6}
\end{equation}

\item Upon receiving $M_3$, \textit{GW} picks up its current timestamp \textit{T$_{gw_1}$} and checks if $|T_{gw_1} - T_{c_1}| < \delta T$.  If the time interval is not within the specified allowable $\delta T$,  $M_3$ is discarded, otherwise, further processing is carried out.  
At this point, \textit{GW} generates a random number \textit{R$_{gw}$}, and calculates a hash \textit{C$_2$} by concatenating \textit{ID$_{GW}$}, $\lambda_{GW}$, and \textit{R$_{gw}$}, as shown in Eq. \ref{eq7}. A message $M_4$=\textit{$\{C_1, C_2, P_{ID_{c}},  T_{gw_1} \}$} is created, XOR with $P_{ID_{GW}}$, i.e., \textit{M$_4$}$ \oplus$ $P_{ID_{GW}}$, and the encrypted message is sent to  \textit{S}, as an authentication request.

\begin{equation}
	C_2 \leftarrow h(ID_{GW} ||\lambda_{GW}||R_{gw}).
	\label{eq7}
\end{equation}

\item Upon receiving $M_4$, \textit{S} picks up its current timestamp \textit{T$_s$}, and checks if $|T_s -  T_{gw_1}| < \delta T$. If the time interval is not within the specified allowable $\delta T$, then $M_4$ is discarded, otherwise, further processing is carried out. Once $M_4$ is validated for time, then \textit{S} checks for the validity of \textit{C} and \textit{GW} in its database. 
If it finds the tuples (\textit{$P_{ID_{c}}$} , $\lambda_C$) for \textit{C} and (\textit{$P_{ID_{GW}}$}, $\lambda_{GW}$) for \textit{GW} in its database, the nodes were previously paired. \textit{S} recalculates the hash \textit{C$_{1}^{'}$} and checks if it matches the hash \textit{C$_1$} received from \textit{C}. 
If the same hash was calculated at \textit{S}, \textit{C} is a registered client
and further processing can take place. Similarly, \textit{S} recalculates the hash \textit{C$_2^{'}$}, and checks if it matches the hash \textit{C$_2$} received from \textit{GW}. If there is a match, \textit{GW} is genuine and further processing can take place. In either case, a mismatch signifies that the request is received from an illegitimate gateway and will be ignored.

\item After the validation of \textit{C} and \textit{GW}, \textit{S} calculates a new pseudo-identity $P_{_C}^{New}$ for \textit{C} by generating the hash of \textit{ID$_C$}, $\lambda_C$, \textit{R$_c$}, \textit{T$_{c_1}$} and \textit{T$_s$}, as shown in Eq. \ref{eqa}. Here,  $P_{_C}^{New}$  serves as a new pseudo-identity for \textit{C}, i.e., $P_{ID_{c}}$.

\begin{equation}
P_{ID_{c}} \leftarrow   P_{_C}^{New}  \leftarrow h(ID_C ||\lambda_C||R_c||T_{c_1}||T_s), 
\label{eqa}
\end{equation}    


\textit{S} also calculates a new pseudo-identity $P_{_{GW}}^{New}$ for \textit{GW} by generating the hash of \textit{ID$_{GW}$}, $\lambda_{GW}$ \textit{R$_{gw}$}, \textit{T$_{gw_1}$} and \textit{T$_s$}, as shown in Eq. \ref{eqb}. Here, $P_{_{GW}}^{New}$ serves as a new pseudo-identity for \textit{GW}, i.e., $P_{ID_{GW}}$.

\begin{equation}
 P_{ID_{GW}} \leftarrow  P_{_{GW}}^{New} \leftarrow h(ID_{GW} ||\lambda_{GW}||R_{gw}||T_{gw_1}||T_s).
 \label{eqb}
\end{equation}


\item	Next, \textit{S} generates a series of hash functions and a session key \textit{K$_S$}  for \textit{GW}, as shown in Eq. \ref{eq8}.  K$_S$ contains all the security primitives intended for \textit{GW} because the whole of data exchange between \textit{C} and \textit{S} will transit via \textit{GW}.
A hash \textit{C$_3$} is generated by concatenating \textit{ID$_C$},  \textit{T$_{c_1}$},  and \textit{T$_s$}.
 A hash \textit{C$_4$} is generated by concatenating \textit{K$_S$}, \textit{C$_3$}, and $ P_{ID_{GW}}$. 
Finally, a   hash \textit{C$_5$} is generated by concatenating \textit{C$_3$} and $R_{c}$.

\begin{subequations}
\begin{equation}
  C_3 \leftarrow h(ID_C ||{T_{c_1}}||T_s), 
\label{eq8a}
\end{equation}    
\begin{equation}
K_S \leftarrow h(ID_{GW} ||\lambda_{GW}||R_{gw}|| P_{ID_{GW}}||T_s),
 \label{server key}
\end{equation}
\begin{equation}
C_{4} \leftarrow h(K_S||C_3|| P_{ID_{GW}}),
 \label{eq8d}
\end{equation}
\begin{equation}
C_{5} \leftarrow h(C_3||{R_c}).
 \label{eq8e}
\end{equation}
\label{eq8}
\end{subequations}

\textit{S} generates $M_5$=\textit{$\{T_s, C_3, C_4, C_5\}$}  and broadcast to \textit{GW}.  The generation of different hash functions in Eq. \ref{eq8} makes \textit{M$_5$} extremely difficult for adversaries to crack. Moreover, these hash functions make it extremely difficult to predict \textit{K$_S$} in \textit{M$_5$}.

\item	Upon receiving $M_5$, \textit{GW} picks up its current time stamp \textit{T$_{gw_2}$} and checks if $|T_s - T_{gw_2}| < \delta  T$. If the time interval is not within the specified allowable $\delta T$, then $M_5$ is discarded, otherwise, further processing is carried out. After the validation of $M_5$,   \textit{GW} calculates its new pseudo-identity \textit{P$_{ID_{_{GW}}}$} by generating the hash of  \textit{ID$_{GW}$}, $\lambda_{GW}$, \textit{R$_{gw}$}, \textit{T$_{gw_1}$} and \textit{T$_s$}, as shown in Eq. \ref{eq10a}. 
In this case, \textit{P$_{ID_{_{GW}}}$} serves as the new pseudo-identity for \textit{GW}, similar to the one generated by \textit{S}.

\begin{subequations}
\begin{equation}
  P_{ID_{_{GW}}} \leftarrow h(ID_{GW} ||\lambda_{GW}||R_{gw}||T_{gw_1}||T_s), 
\label{eq10a}
\end{equation}    
\begin{equation}
  K_{GW} \leftarrow h(P_{ID_{c}}||ID_{GW}||R_c||R_{gw}), 
\label{eq11a}
\end{equation}    
\begin{equation}
C_6 \leftarrow h(K_{GW}||{C_3}).
 \label{eq11b}
\end{equation}
\end{subequations}

\item	Next, a session key \textit{K$_{GW}$}  is generated by \textit{GW}  using a hash function to concatenate  $P_{ID_{c}}$, \textit{ID$_{GW}$}, \textit{R$_c$} and \textit{R$_{gw}$}, as shown in Eq. \ref{eq11a}. Finally, a 
hash \textit{C$_6$} is calculated by concatenating \textit{K$_{GW}$} with \textit{C$_3$}, as shown in Eq. \ref{eq11b}.

At this point, \textit{GW} creates a message $M_6$=\textit{$\{ C_{{5}}, C_6, T_s, T_{gw_2} \}$}, XOR with $P_{ID_{_{GW}}}$, i.e., \textit{M$_6$}$ \oplus$ $P_{ID_{_{GW}}}$, and the encrypted message
is broadcast to \textit{C}.

\item	Upon receiving $M_6$, \textit{C} picks up its current time stamp \textit{T$_{c_2}$} and checks if $|T_{c_2} - T_{gw_2}| < \delta  T$. If the time interval is not within the specified allowable $\delta T$, then $M_6$ is discarded, otherwise, further processing is carried out. 
After the validation of $M_6$, \textit{C} calculates its new pseudo-identity $P_{ID_{_{c}}}^{New}$ by generating the hash of \textit{ID$_C$}, $\lambda_C$, \textit{R$_c$}, \textit{T$_{c_1}$} and \textit{$T_s$}, as shown in Eq. \ref{eq12a}. In this case, $P_{ID_{_{c}}}^{New}$ serves as a new pseudo-identity for \textit{C}. It is worth mentioning that the same pseudo-identity for \textit{C} was generated earlier by {\textit{S}}.

\begin{subequations}
\begin{equation}
P_{ID_{_{c}}}^{New} \leftarrow h(ID_C ||\lambda_C||R_c||T_{c_1}||T_s), 
\label{eq12a}
\end{equation}    
\begin{equation}
K_{C} \leftarrow h(P_{ID_{gw}} ||ID_C||R_{c}||R_{gw}).
 \label{eq12c}
\end{equation}
\end{subequations}

It then recalculates the hash \textit{C$_5$}. 
{If \textit{$P_{ID_{_{c}}}^{New}$} holds, i.e., $P_{ID_{_{c}}}^{New}$ == \textit{h($\lambda_C ||  C_{5}$)},}   
the identity of \textit{GW} is successfully verified.  
At this point, \textit{C} generates a session key \textit{K$_C$} based on a hash function and concatenating P$_{ID_{_{gw}}}$, ID$_C$, R$_c$ and R$_{gw}$, as shown in Eq. \ref{eq12c}. To check the validity of \textit{K$_C$}, \textit{C} recalculates \textit{C$_6$}, i.e., \textit{C$^{'}_{6}$}, which is calculated as  \textit{h}($P_{ID_{_{c}}} || ID_{GW} || R_c || R_{gw} || C_3$). It can also be calculated as \textit{h}(K$_{GW}||C_3$.
If \textit{C$^{'}_{6}$} matches the \textit{C$_6$} received from \textit{GW}, i.e., $C_6$ == $h(K_C ||T_{c_1})$, \textit{K$_C$} is valid.

Going through this process successfully, \textit{C} and \textit{GW} have mutually authenticated each other and are authorized to transmit the healthcare data to \textit{S}. The overall authentication process is shown in Fig. \ref{fig2}.

\end{enumerate}

\begin{figure*}
\centering
  \includegraphics[width=.86\linewidth, height = 14.42 cm]{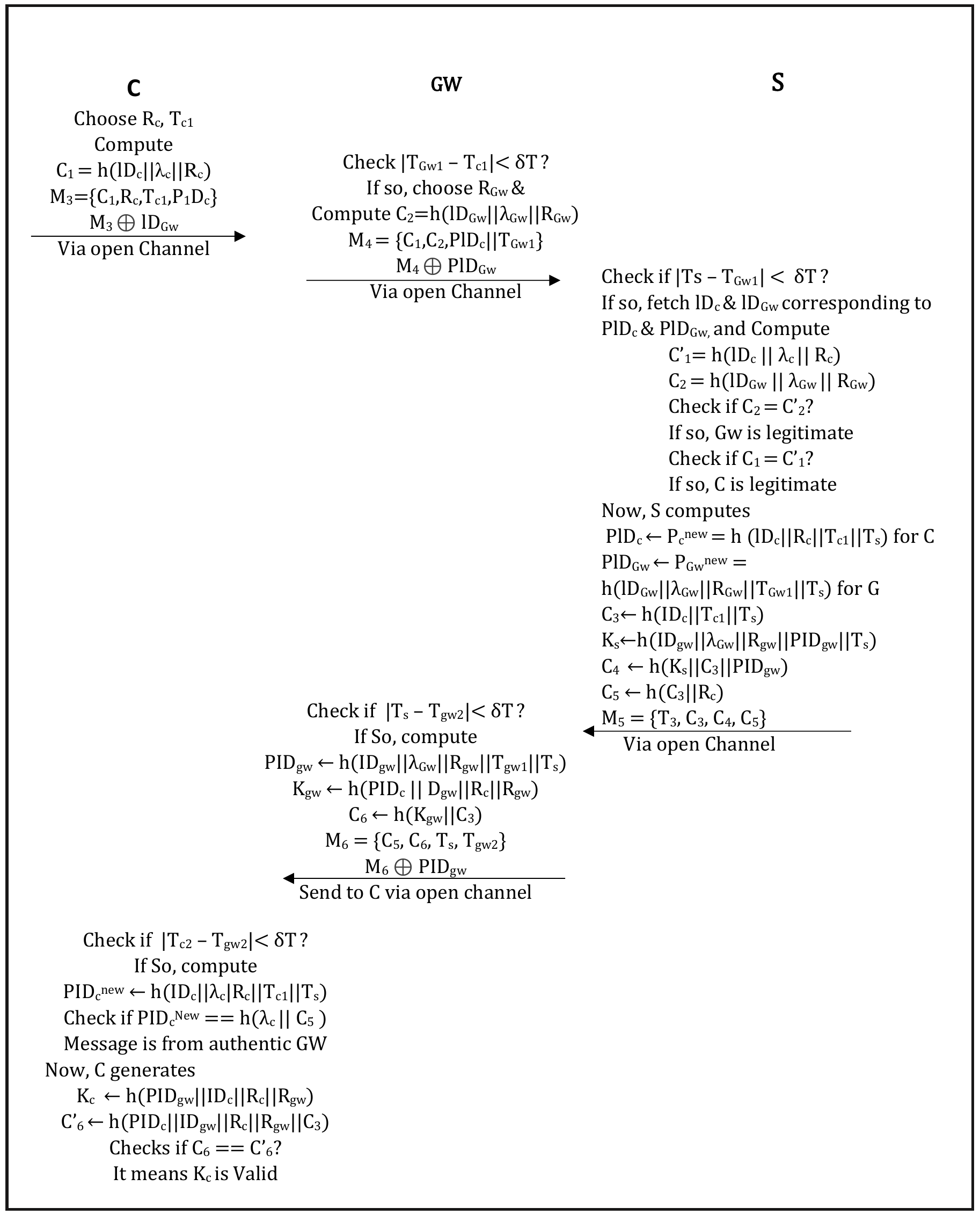}
  \caption{Authentication Phase}
  \label{fig2}
\end{figure*}

\section{Security Analysis} \label{section5}
To check the validity of the \sol design, informal analysis is conducted, which shows that \sol is resilient against a number of adversarial attacks. In \sol, IDs and pseudo-random numbers are 128 bit, and timestamps are 32 bit in length. We used the
SHA3-256 hash function to generate a hash digest of 256-bit length. 

\subsection{Replay Attack}
A timestamp is used in each message by \textit{C}, \textit{GW} and \textit{S} to protect the content of these messages from replay attack. 
Therefore, the validity of each message can be checked. If a message is not within the legal time delay, i.e., $\delta t$, it will be discarded. Similar to \sol, the existing schemes are resilient to replay attacks. 

\subsection{Forgery Attack } 
An adversary can launch a forgery attack on all the network entities (clients, servers, and gateways). We discuss all the possible scenarios below.

\subsection{Forgery Attack on Server}

If an adversary launches a forgery attack on \textit{S}, it will need to capture and manipulate 
\textit{M$_1$} and \textit{M$_4$}. For \textit{M$_4$}, the adversary needs to provide a valid \textit{C$_1$}, \textit{C$_2$} and \textit{P$_{ID_{c}}$} to \textit{S}. Due to the encrypted $M_4$ (\textit{M$_4$}$\oplus$ $P_{ID_{GW}}$), the adversary will initially require \textit{P$_{ID_{GW}}$} to extract the content (\textit{C$_1$}, \textit{C$_2$} and \textit{P$_{ID_{c}}$}).  Even if it acquires  \textit{P$_{ID_{GW}}$}, it will require $\lambda_C$ to crack \textit{C$_1$} and  \textit{ID$_{GW}$} to crack $C_2$. An adversary may also try to eavesdrop and manipulate \textit{M$_1$} to launch a forgery attack on \textit{S}.  However, due to the encrypted \textit{M$_1$}$ \oplus$ $P_{ID_{c}}$, the adversary will require  $P_{ID_{c}}$ to crack this message. Even if it cracks it, the adversary would still need $\lambda_C $ and \textit{r$_c$} to regenerate a valid \textit{D$_1$}. The use of hash functions, pseudo-random numbers and secret keys makes it extremely difficult to launch forgery attacks on \textit{S}. 

\subsection{Forgery Attack on Client}
To launch a forgery attack on \textit{C}, an adversary  will need to capture and manipulate 
\textit{M$_2$} and \textit{M$_6$}. To forge \textit{M$_2$}, a valid \textit{D$_2$} needs to be presented to \textit{C}. To do so, the adversary would require $P_{ID_{c}}$ of \textit{C} and $ID_{GW}$ of \textit{GW}. For \textit{M$_6$}, $P_{ID_{_{GW}}}$ is required to decrypt it. Even if an adversary decrypts \textit{M$_6$}, the former will require to crack $C_5$ and $C_6$ to launch a forgery attack on \textit{C}. In \sol, \textit{C$_5$} and \textit{C$_6$} are the most resilient and robust hashes as they are composed of secret keys and pseudo-random numbers. The keys themselves are hashed making it highly unlikely for an adversary to crack them even with the most sophisticated hardware and software platforms. For a successful forgery attack on \textit{C}, an adversary needs to know these hashes, keys, and pseudo-random numbers.

\subsection{Forgery Attack on Gateway}

To launch a forgery attack on \textit{GW}, an adversary needs to capture and manipulate 
\textit{M$_3$} and \textit{M$_5$}. 
To forge \textit{M$_3$}, an adversary would initially require a valid $ID_{GW}$ to crack it. Even if it cracks it, a valid \textit{C$_1$} and $P_{ID_c}$ need to be presented to \textit{GW}. For 
 $M_5$, a number of hash functions ($C_3$, $C_4$, and $C_5$) are required. The complex combination of these hash functions in \textit{M$_5$} makes the latter extremely difficult for adversaries to decrypt. Moreover, these hash functions make it extremely difficult to predict \textit{K$_S$} in \textit{M$_5$}.

\subsection{Untraceability of Client and Gateway}
\textcolor{black}{Both \textit{C} and \textit{GW} get a new pseudo-identity in every new session and these pseudo-identities are always different from previous ones due to their unique timestamps. Therefore, \textit{C} and \textit{GW} are untraceable because their real/actual identities are never disclosed in the exchanged messages.}

\subsection{Mutual Authentication and Key Agreement}
\textcolor{black}{Mutual authentication is guaranteed because none of the entities of any session can be forged. Every session is managed under a unique session key to encrypt the information exchanged during a session.}

\subsection{De-Synchronization Attack }
\textcolor{black}{If \textit{M$_2$} is not received by \textit{C} during the pairing phase due to network delays or blockage by an adversary, then \textit{GW} can continue its operations according to the last updated values for next pairing. If \textit{M$_5$} is not received by \textit{GW} in the authentication phase due to network delays or blockage by an adversary, \textit{GW} can also continue its operations according to the last updated values for next session. If \textit{M$_6$} is not received by \textit{C} in the authentication phase due to network delays or blockage by an adversary, \textit{GW} can use the latest $P_{ID_{c}}$ to complete the process.}

\subsection{Availability}

\textcolor{black}{The majority of existing schemes use long-term keys at the beginning and maintain them for pairing and authentication. However, in \sol, there are no long-term keys for \textit{C} and \textit{GW}. The pairing phase is mandatory for every new \textit{C} and \textit{GW}, since they are not supposed to have any information about each other before their initial interaction.}

\begin{table*}[ht]
\centering
\caption{Resilience against various Attacks}
\vspace{-0.3cm}
\label{resilence against attacks}
\begin{tabular}{p{4.7cm}p{1.5cm}p{1.5cm}p{1.5cm}p{1.6cm} p{1.6cm} }
\toprule
\multicolumn{1}{l}{\textit{Attacks}} & \cite{amin2016secure}  &\cite{li2017anonymous} & \cite{jan2018sams}                                                                                                                                                                                                        & \cite{gope2016realistic} &\textit{\sol} 
 \\ \midrule[\heavyrulewidth]        
Replay                 & No & Yes  & Yes &  Yes & Yes \\ 
Resistance to Server Forgery   & No & Yes & No & No & italic\\
Resistance to Client Forgery    & Yes & No & No & Yes & Yes\\
Resistance to Gateway Forgery   & Yes & No & No & Yes & Yes\\

Untraceability   & No & Yes  & No & No & Yes\\
Mutual Authentication   & Yes & Yes  & Yes & Yes & Yes\\

Key Agreement   & Yes & Yes  & Yes & Yes & Yes\\
De-Synchronization & No & No & No & No & Yes
\\

Availability & No & No & No & No & Yes
\\
 \midrule[\heavyrulewidth]
\end{tabular}
\end{table*}

\section{Performance Evaluation} \label{simulation}
In this section, we evaluate and validate our proposed approach through experimental results in a simulation environment. In addition, we used NS-2 as a simulation tool to implement and validate different protocols. Initially, the network infrastructure is developed through the random deployment of sensor devices, gateways,  and remote servers. To evaluate the efficiency of our scheme, we increased the number of sensor devices, gateways and remote servers in the deployed area followed by an increase in the network traffic. To highlight its efficiency, we compare \sol against existing approaches in terms of computational and communication overhead, individual device lifetime statistics followed by network lifespan, and latency.

\subsection{Computation Overhead}

In Table \ref{Table 1}, we provide a summary of the computational overhead comparison against the evaluated schemes. In this table, $T_h$ and $T_{XOR}$ refer to the computational time needed to perform the hash and XOR operations at \textit{C}, \textit{GW} and \textit{S}, respectively. In \cite{li2017anonymous}, the gateways do not perform any computation. Instead, they forward the messages directly to a hub, i.e., a server. As a result, the computational overhead at the gateway is left blank. Among the existing schemes, \cite{amin2016secure} incurs relatively higher computational overhead in comparison to \cite{jan2018sams}, \cite{gope2016realistic} and \cite{li2017anonymous}. The comparison in this table highlights the effectiveness of \sol as it generates highly secure and composite hash functions with the least computational overhead. More importantly, the relatively smaller computational overhead is incurred at resource-constrained wearables, which makes \sol a feasible option for deployment in large-scale healthcare applications.

\begin{table*}[h]
\centering

\caption{Computation Overhead Comparison}
\vspace{-0.3cm}
\begin{tabular}{p{3.2cm}p{2.44cm}p{2.44cm}p{2.5cm}p{3cm}}
\hline
\toprule
Schemes            & Client (T$_C$) &Gateway (T$_G$)&   Server (\textit{T$_S$}) & Total Cost \\
\hline
Amin et. al \cite{amin2016secure}   & $5T_h$+$3T_{XOR}$             &$12T_h$+$7T_{XOR}$           &   $15T_h$+$7T_{XOR}$         & $32T_h$+$17T_{XOR}$                       \\
Li et. al \cite{li2017anonymous}   & $13T_h$+$7T_{XOR}$         &\qquad -       &   $4T_h$+$12T_{XOR}$         & $17T_h$+$19T_{XOR}$                       \\
Jan et. al \cite{jan2018sams}   & $6T_h$+$1T_{XOR}$             &$7T_h$+$1T_{XOR}$           &   $10T_h$+$2T_{XOR}$         & $23T_h$+$4T_{XOR}$                       \\
Gope et. al \cite{gope2016realistic}   & $3T_h$+$1T_{XOR}$  &$14T_h$+$7T_{XOR}$                       &   $9T_h$+$4T_{XOR}$         & $26T_h$+$12T_{XOR}$                       \\
\sol   & $5T_h$+$2T_{XOR}$    &$4T_h$+$2T_{XOR}$                    &   $8T_h$+$1T_{XOR}$         & $17T_h$+$5T_{XOR}$                       \\
\bottomrule
\label{Table 1}
\end{tabular}
\end{table*}

\subsection{Communication Overhead}

In Table \ref{Table 2:communication cost}, we show the communication overhead incurred by the network entities while exchanging the messages among themselves. In \sol, the encrypted message (\textit{M$_1$}$ \oplus$ $P_{ID_{c}}$) is transmitted by \textit{C}. \textit{M$_1$} has a length of $544$ bits.
The message \textit{M$_2$}  is transmitted by \textit{S}  as $M_2$=\textit{$\{D_2, t_s \}$} and the communication overhead incurred is $288$ bits.
The message \textit{M$_3$} is transmitted by \textit{C} as $M_3$=\textit{$\{C_1, R_c, T_{c_1}, P_{ID_c} \}$} and has a length of $544$ bits.
The encrypted \textit{M$_4$} (\textit{M$_4$}$ \oplus$ $P_{ID_{GW}}$) incurs a communication overhead of $672$ bit on \textit{GW}. The most sophisticated and complex \textit{M$_5$}=\textit{$\{T_s, C_3, C_4, C_5\}$} has multiple hash functions and incurs a communication overhead of 800 bits on \textit{S}.
Finally, the encrypted  \textit{M$_6$} (\textit{M$_6$}$ \oplus$ $P_{ID_{_{GW}}}$) incurs a communication overhead of $576$ bits on \textit{GW}. The total communication overhead incurred by network entities in our proposed approach is $3424$ bits. In comparison to the existing schemes of \cite{amin2016secure}, \cite{li2017anonymous} and \cite{jan2018sams}, \sol has a lower communication overhead, but it has relatively higher overhead compared to \cite{gope2016realistic}. However, this comparison does not signify that \cite{gope2016realistic} is superior to \sol in terms of communication overhead. In any scheme for resource-constrained wearable devices, the overhead imposed on wearables themselves is the most important factor. To this end, \sol incurs a communication overhead of $1088$ bits on a wearable device compared to $1340$ bits of \cite{gope2016realistic}.




\begin{table}[h]
\centering
\caption{Communication Overhead Comparison}
\vspace{-0.3cm}
\begin{tabular}{p{2cm}p{2.5cm}p{2.5cm}}
\hline
\toprule
Schemes            & Number of Messages & Number of Bits\\
\hline
Amin et. al \cite{amin2016secure}   & \qquad $6$ &\qquad $4096$ \\
Li et. al \cite{li2017anonymous}   & \qquad $4$       &\qquad $4672$  \\
Jan et. al \cite{jan2018sams}   & \qquad $5$ &\qquad $3808$           \\
Gope et. al \cite{gope2016realistic}   & \qquad $4$  &\qquad $3184$    \\
\sol   & \qquad $6$    &\qquad $3424$                            \\
\bottomrule
\label{Table 2:communication cost}
\end{tabular}
\end{table}
\subsection{Network Lifespan Analysis Against Field-Proven Schemes}

The performance reliability of any authentication scheme is dependent on the network lifetime. Therefore, network lifespan needs special attention while designing a new authentication scheme for resource-limited networks. Keeping in mind the reliability factor of an authentication scheme, we evaluate \sol in terms of the lifetime of individual sensor devices and the whole network lifespan in comparison to existing schemes. The simple authentication with accurate results of \sol is effective in terms of network lifetime because the legitimate devices need only two messages to verify the legitimacy of communicating devices. Besides that, the simple authentication process of \sol with the least computation and communication costs minimizes the energy consumption during handshake among the participating devices. During simulations, \sol showed superior results of individual device lifetime and network lifespan, due to its light computation and storage overhead on wearable devices. 
Figures \ref{f1} and \ref{f2} present the results of \sol along with existing state-of-the-art schemes for individual device lifetime and network lifetime.

\subsection{End-to-End Delay Analysis}

\begin{figure*}[htbp]
  	\begin{minipage}[t]{0.32\linewidth}
  		\includegraphics[width=\linewidth]{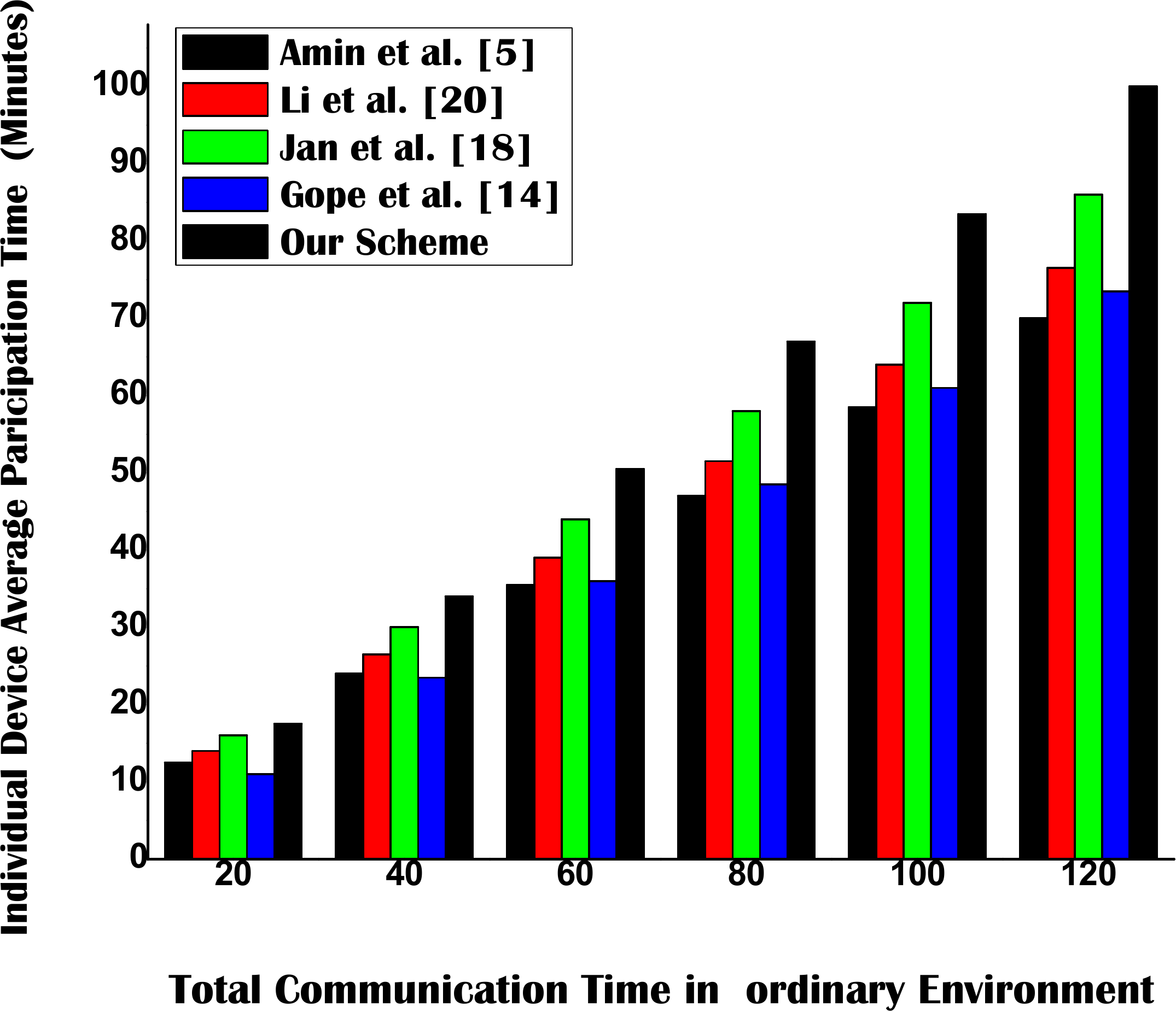}
  		\caption{Individual device lifespan results.}
  		\label{f1}
  	\end{minipage}%
  	\hfill%
  	\begin{minipage}[t]{0.32\linewidth}
  		\includegraphics[width=\linewidth]{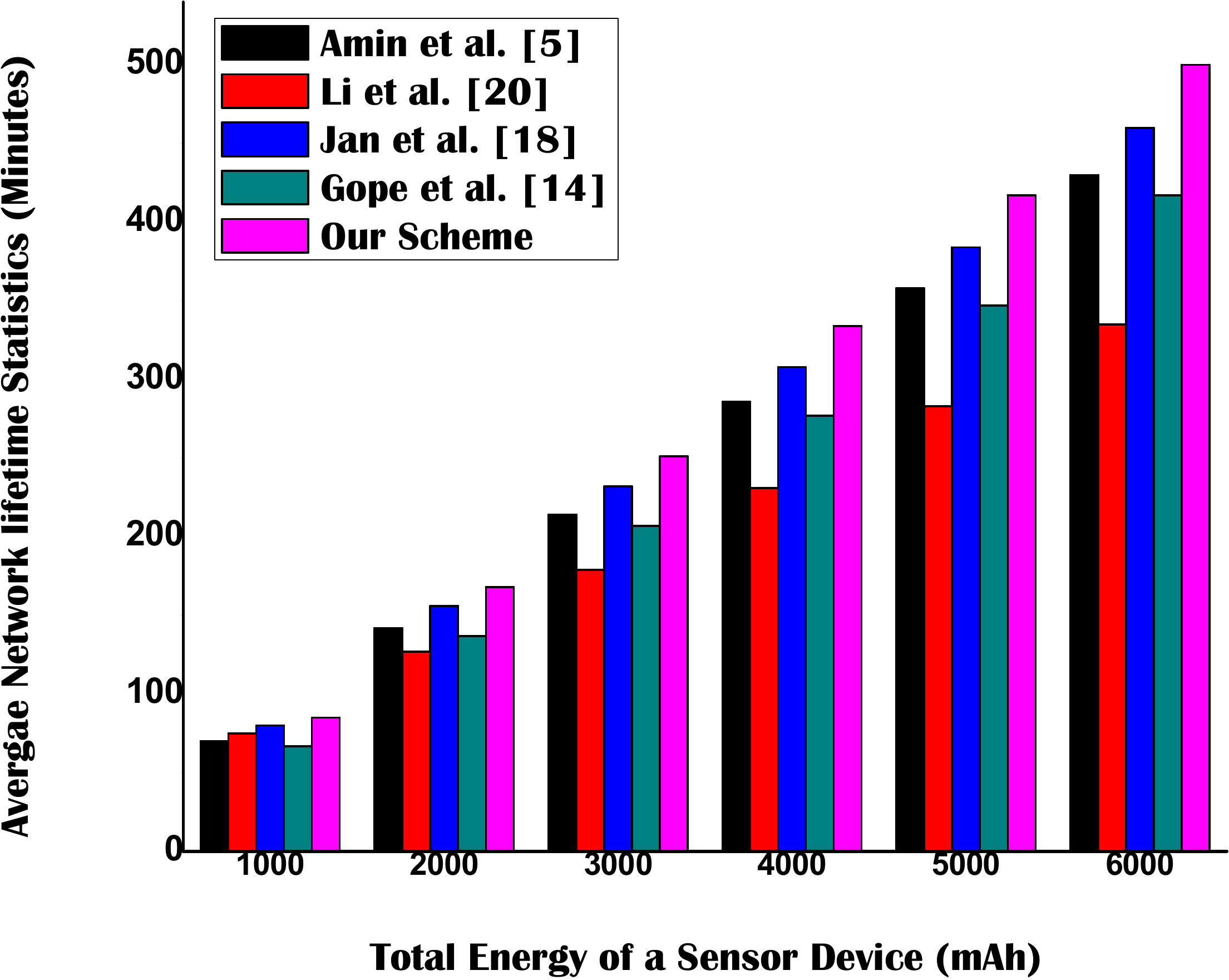}
  		\caption{Network lifespan results. }
  		\label{f2}
  	\end{minipage} 
  	\hfill%
  	\begin{minipage}[t]{0.32\linewidth}
  		\includegraphics[width=\linewidth]{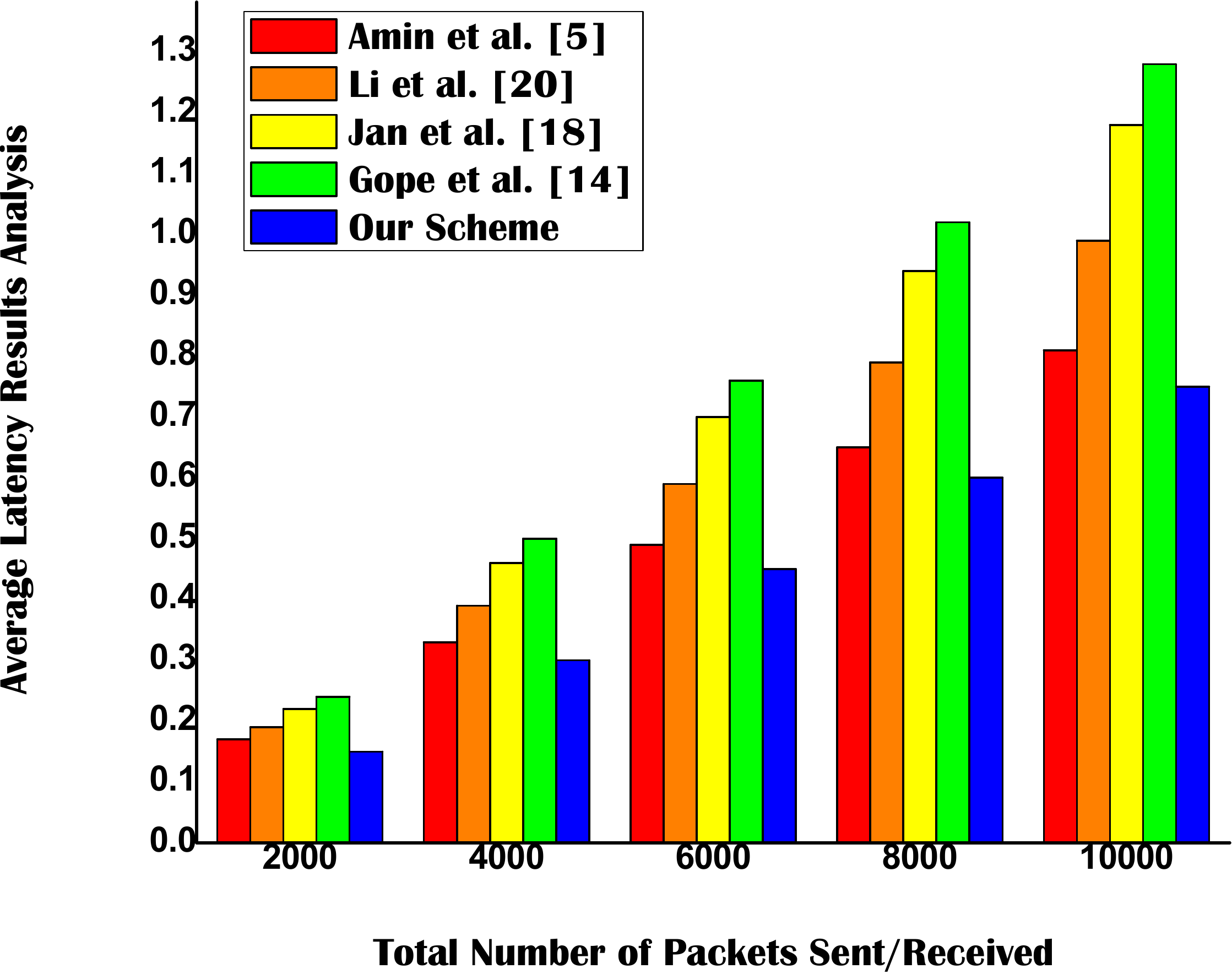}
  		\caption{Latency results.}
  		\label{f3}
  	\end{minipage}%
  	\hfill%
  \end{figure*}

In delay-sensitive applications of IoT networks, the performance of any protocol is dependent on latency, since additional delay in the deployed network disrupts its effectiveness. To this end, we have evaluated the latency of \sol. The simple authentication process and lightweight nature of \sol ensure its efficiency, while the time consistency observed during the communication process was noteworthy. Furthermore, we have increased network traffic with the addition of new devices in the simulation environment. However, during the communication process, the transmission and reception of messages showed a constant time frame throughout the entire process. 
Our results presented in Figure~\ref{f3} demonstrate that \sol incurs significantly lower latency in comparison to state-of-the-art schemes.


\section{Conclusion}\label{conclusion}
In this paper, we proposed \sol, a lightweight yet highly secure scheme for green communications, focusing on biomedical data in IoT-enabled health informatics. \sol has three phases that facilitate the resource-constrained wearable devices to initiate simple registration and authentication procedures with a mobile gateway and a remote server. The registration requires two messages to register the wearables with a remote server and the authentication relies on four such messages to establish a secure end-to-end session for data exchange among the communicating entities. \sol uses lightweight hash functions and XOR operations to accomplish these phases and is highly efficient for the immediate delivery of time-critical and delay-sensitive data. The experimental results verify the efficiency of \sol, as it is highly resilient against a number of attack scenarios and, at the same time, incurs low computational and communication overheads. 
\textcolor{black}{The limitation of LightIoT is the validation of mobile wearable devices in an operational environment, because the one step registration process is performed in the offline phase.}\\

\ifCLASSOPTIONcaptionsoff
  \newpage
\fi

\section*{Acknowledgments}
This work is partially supported by the NIH (P20GM109090), NSF (CNS-2016714), and the Nebraska University Collaboration Initiative.

\bibliographystyle{IEEEtran}
\bibliography{bibliography}

\section*{Biographies}

\vskip -4.0\baselineskip plus -1fil

\begin{IEEEbiography}[{\includegraphics[width=1in,height=1.25in,clip,keepaspectratio]{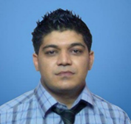}}]{ Mian Ahmad Jan}  is an assistant professor at the department of computer science, Abdul Wali Khan University Mardan, Pakistan. He completed his PhD in Computer Systems at the University of Technology Sydney (UTS), Australia in 2016. He had been the recipient of various prestigious scholarships during his PhD studies. He was the recipient of International Research Scholarship (IRS), UTS and Commonwealth Scientific Industrial Research Organization (CSIRO) scholarships.  His research interests include energy-efficient and secured communication in Wireless Sensor Networks, Internet of Things, Edge Computing.   His research has been published in prestigious IEEE Transactions and core-ranked conferences. He has been guest editor of numerous special issues in various prestigious journals such as IEEE Transactions on Industrial Information, Springer Neural Networks and Applications, and Elsevier Future Generation Computer Systems etc. 
\end{IEEEbiography}
\vskip -2.0\baselineskip plus -1fil

\begin{IEEEbiography}[{\includegraphics[width=1in,height=1.25in,clip,keepaspectratio]{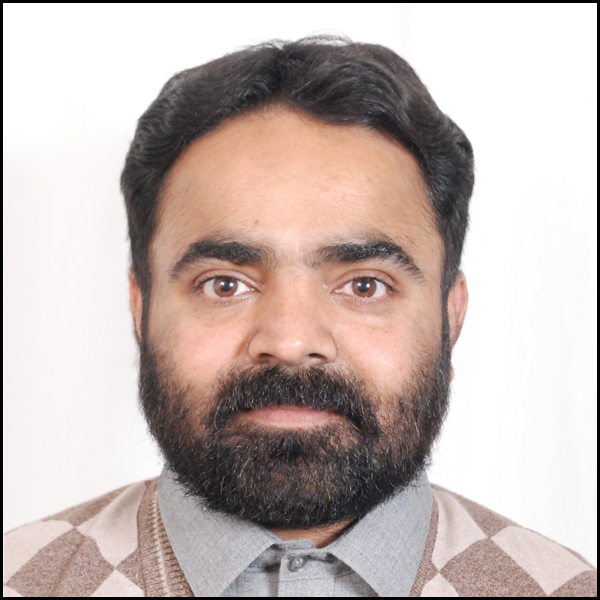}}]{ Fazlullah Khan} (Senior Member IEEE), is an Assistant professor of Computer Science at Abdul Wali Khan University Mardan (AWKUM), Pakistan. He had been the recipient of various prestigious scholarships during his PhD studies and has been awarded the best researcher awarded for the year 2017. His research interests are Intelligent and robust protocol designs, Security and Privacy of Wireless Communication Systems, Internet of Things, Machine Learning, Artificial Intelligence. Recently, he has been involved in latest developments in the field of Internet of Vehicles security and privacy issues, Software-defined Networks, Fog Computing and Big Data Analytics. He has published his research work in top-notch journals and conferences. His research has been published in IEEE Transactions on Industrial Informatics, IEEE Internet of Things, IEEE Access, Elsevier Computer Networks, Elsevier Future Generations Computer Systems, Elsevier Journal of Network and Computer Applications, Elsevier Computers and Electrical Engineering, Springer Mobile Networks and Applications. 
\end{IEEEbiography}

\vskip -2.0\baselineskip plus -1fil

\begin{IEEEbiography}[{\includegraphics[width=1in,height=1.25in,clip,keepaspectratio]{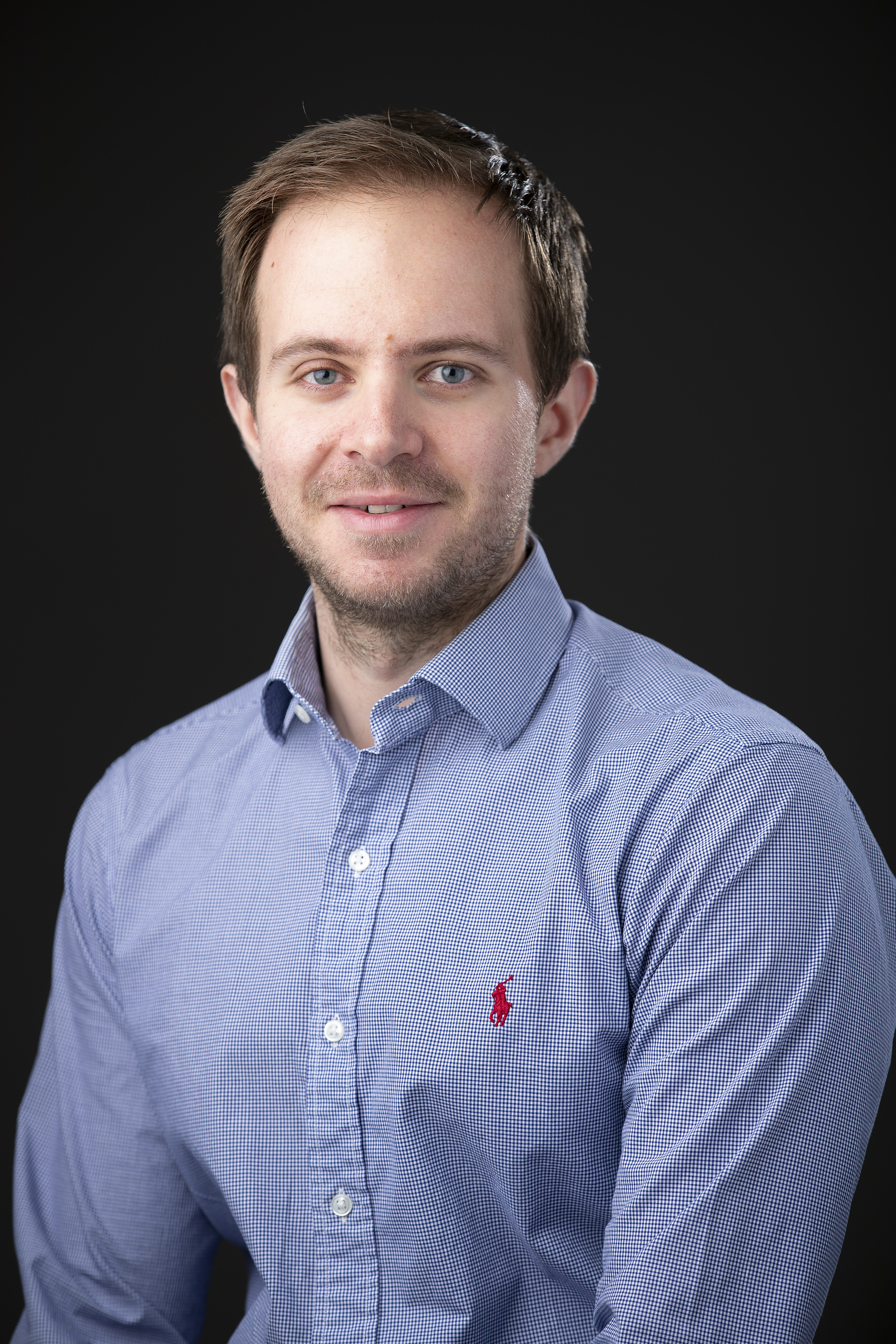}}]{Spyridon Mastorakis} is an Assistant Professor in Computer Science at the University of Nebraska Omaha. He received his Ph.D. in Computer Science from the University of California, Los Angeles (UCLA) in 2019. He also received an MS in Computer Science from UCLA in 2017 and a 5-year diploma (equivalent to M.Eng.) in Electrical and Computer Engineering from the National Technical University of Athens (NTUA) in 2014. His research interests include network systems and protocols, Internet architectures, IoT and edge computing, and security.
\end{IEEEbiography}

\vskip -2.0\baselineskip plus -1fil

\begin{IEEEbiography}[{\includegraphics[width=1in,height=1.25in,clip,keepaspectratio]{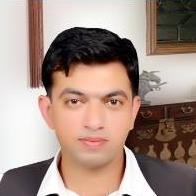}}]{\textbf{Muhammad Adil}} received his Associate Engineer degree in Electronics, BS degree in computer science and MS (CS) degree with specialization in Computer Networks from Virtual University of Pakistan, Lahore in 2016 and 2019, respectively. He has CCNA and CCNP certification. He is currently a PhD student. His research area includes different routing protocols, Security, and Load Balancing in WSN and IoT networks. Moreover, Mr. Adil is also interested	in Dynamic Wireless Charging of Electric Vehicles connected in network	topological infrastructure with Machine learning techniques. He has many	publications in prestigious journals such as IEEE Internet of Things, IEEE Access, Computer Networks Elsevier,	MDPI Sensor, CMC-Computer Material \& Continua, etc. In addition, he is IEEE Student member, honorary Member of London Journals (Press London Journal of Research in Computer Science and Technology (LJRCST)) and European Alliance for Innovation (EAI). He is reviewing for prestigious journals, such as IEEE Access, IEEE Sensors, IEEE Systems, IEEE Internet of Things, MDPI Sensors and Computer Networks Elsevier Journals.
\end{IEEEbiography}

\vskip -2.0\baselineskip plus -1fil

\begin{IEEEbiography}[{\includegraphics[width=1in,height=1.25in,clip,keepaspectratio]{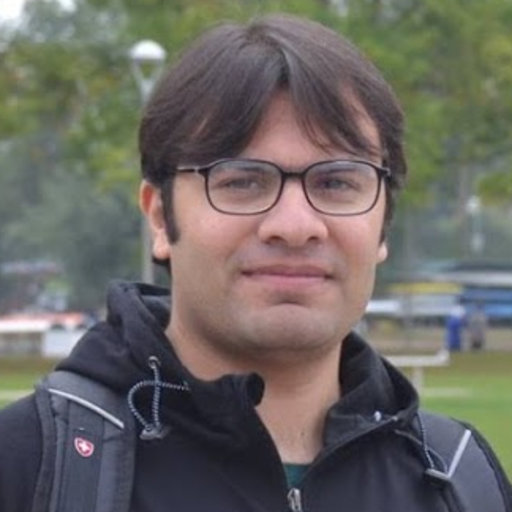}}]{\textbf{ Aamir Akbar}} received Ph.D degree in Computer Science from Aston University (UK) in 2019. He completed an MSc degree at Oxford Brooks University (UK) in 2012. Currently, he is a lecturer at the Department of Computer Science, Abdul Wali Khan University Mardan, Pakistan. His research interests include, but not limited to, AI techniques for IoT, SDN, energy-efficient fog/edge computing. His work leverages multi-objective optimisation, evolutionary computation, self-adaptivity and self-awareness to tackle problems. Also, he has published his work in prestigious IEEE journals and Transactions.
\end{IEEEbiography}

\vskip -2.0\baselineskip plus -1fil

\begin{IEEEbiography}[{\includegraphics[width=1in,height=1.25in,clip,keepaspectratio]{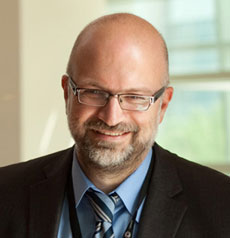}}]{Nicholas Stergiou} is the Distinguished Community Research Chair in Biomechanics and Professor and the Director of the Biomechanics Research Building and the Center for Research in Human Movement Variability at the University of Nebraska at Omaha. Recently he was also appointed as the Assistant Dean and Director of the Division of Biomechanics and Research Development. He is the Founding Chair of the first ever academic Department of Biomechanics that graduates students with a BS in Biomechanics. His research focuses on understanding variability inherent in human movement and he is an international authority in the study of Nonlinear Dynamics. He has published more than 200 peer-reviewed papers and have been inducted to the National Academy of Kinesiology and as a Fellow to the American Institute for Medical and Biological Engineering and the American Society of Biomechanics. He is currently serving as the President of the American Society of Biomechanics.
\end{IEEEbiography}
\end{document}